\documentclass[12pt]{iopart}
\usepackage{amsmath}
\usepackage{amssymb}
\usepackage{color}
\usepackage{graphicx}
\usepackage{physics}
\usepackage{subfig}
\usepackage[hidelinks,colorlinks]{hyperref}
\usepackage[ruled,linesnumbered]{algorithm2e}

\usepackage{biblatex}
\graphicspath{{fig/}}

\def\equationautorefname~#1\null{%
  Eq.~(#1)\null
}

\begin{document}

\title[Random-depth Quantum Amplitude Estimation]{Random-depth Quantum Amplitude Estimation}

\author{Xi Lu}
\address{School of Mathematical Science, Zhejiang University, Hangzhou, 310027, China}
\ead{luxi@zju.edu.cn}

\author{Hongwei Lin}
\address{School of Mathematical Science, Zhejiang University, Hangzhou, 310027, China}
\ead{hwlin@zju.edu.cn}

\vspace{10pt}
\begin{indented}
\item[]Nov. 2023
\end{indented}

\begin{abstract}
    The maximum likelihood amplitude estimation algorithm~(MLAE) is a practical solution to the quantum amplitude estimation problem with Heisenberg limit error convergence.
    We improve MLAE by using random depths to avoid the so-called critical points, and do numerical experiments to show that our algorithm is approximately unbiased compared to the original MLAE and approaches the Heisenberg limit better.
\end{abstract}

%
%
%
%
%


\section{Introduction}\label{sec:intro}

Quantum computing is an emerging subject that studies faster algorithms on quantum computers over classical ones to solve some specific problems.
Early quantum algorithms have achieved astonishing speedups over known classical algorithms, such as the quadratic speedup of Grover's search~\cite{Grover1997}, and the exponential speedup of Shor's integer factorization~\cite{Shor1997}.
Later algorithms like quantum approximate optimization algorithms (QAOA)~\cite{qaoa,zhou2020quantum,farhi2016quantum}, variational quantum eigen solver (VQE)~\cite{peruzzo2014variational,mcclean2016theory} and quantum neural networks (QNN)~\cite{pesah2021absence,schuld2014quest} also shows great potentials in quantum computing.

The \textit{amplitude estimation} problem~\cite{brassard2002quantum} is one of the most fundamental problems in quantum computing.
Let $\mathcal{A}$ be any quantum algorithm that performs the following unitary transformation,
\begin{align}
    \mathcal{A}\ket{00\cdots0} = & \sqrt{1-a}\ket{\psi_0}\ket{0} + \sqrt{a}\ket{\psi_1}\ket{1} \nonumber
    \\ = & \cos\phi\ket{\psi_0}\ket{0} + \sin\phi\ket{\psi_1}\ket{1}.
    \label{eq:def-A}
\end{align}
The goal of amplitude estimation problem is to estimate $a$, with as few oracle calls to $\mathcal{A}$, along with its inverse and controlled-$\mathcal{A}$, as possible.
It is a problem generalized from phase estimation and quantum counting, and has been widely applied in quantum chemistry~\cite{aspuru2005simulated,lanyon2010towards,knill2007optimal}, machine learning~\cite{wiebe2014nearest,wiebe2014deep}, risk analysis~\cite{woerner2019quantum,egger2020credit} and option pricing~\cite{stamatopoulos2020option,rebentrost2018quantum} in recent studies.

The earliest solution~\cite{brassard2002quantum} is a combination of the quantum Fourier transformation (QFT) based phase estimation and Grover's search.
There are some later researches~\cite{Svore2013,kimmel2015robust,Wiebe2016} that improve the robustness of phase estimation.
The paper~\cite{van2023quantum,van2022quantum} first proposes an unbiased amplitude estimation algorithm based on the idea of random phase shift in the unbiased QPE algorithm~\cite{linden2022average}.
Later works on the same topic include~\cite{lu2022unbiased,lu2023unbiased,rall2023amplitude}.
\footnote{We thank Dr. Arjan Cornelissen, one of the author of~\cite{van2022quantum}, for letting us know that the paper~\cite{van2022quantum} was published in arXiv earlier than the paper~\cite{lu2022unbiased}. }
The modified Grover's operator~\cite{uno2021modified} is an approach that is designed to perform robustly under depolarizing noise.

Recent researches also study amplitude estimation algorithms without the use of QFT since it is believed that the the large entanglement brought by QFT could make the algorithm less robust on noise intermediate-scale quantum (NISQ) devices~\cite{suzuki2020amplitude}.
Another reason is that one needs two oracle calls to $\mathcal{A}$ to construct a rotation matrix by angle $2\phi$ in QFT, which brings a double factor in the number of oracle calls.
The \textit{maximum likelihood amplitude estimation} (MLAE)~\cite{suzuki2020amplitude} algorithm is an approach without QFT, which is proved to have an error convergence $O(N^{-1})$ asymptotically when using an exponential incremental sequence (EIS), which is quadratically faster than $O(N^{-1/2})$ for classical Monte Carlo algorithm.
The error convergence $O(N^{-1})$ is also known as the Heisenberg limit~\cite{dutkiewicz2022heisenberg}.
There is a variant of MLAE~\cite{tanaka2022noisy} that is built for noisy devices without estimating the noise parameters.
The depth-jittering quantum amplitude estimation (DJQAE)~\cite{callison2022improved} improves MLAE by jittering the Grover depth to avoid the so-called exceptional points of MLAE.
The iterative quantum amplitude estimation (IQAE)~\cite{Grinko2021} is another approach without phase estimation by iteratively narrows the confidence interval of amplitude, which is proved rigorously to achieve a quadratic speedup up to a double-logarithmic factor compared to classical Monte Carlo (MC) estimation.
The iterative quantum phase estimation protocol for shallow circuits~\cite{smith2022iterative} is a two-step protocol for near-term phase estimation that also avoids the use of QFT.
There are also several other approaches~\cite{aaronson2020quantum,nakaji2020faster,zhao2022adaptive,Plekhanov2022variationalquantum}.

\begin{figure}
  \centering
  \includegraphics[width=.5\textwidth]{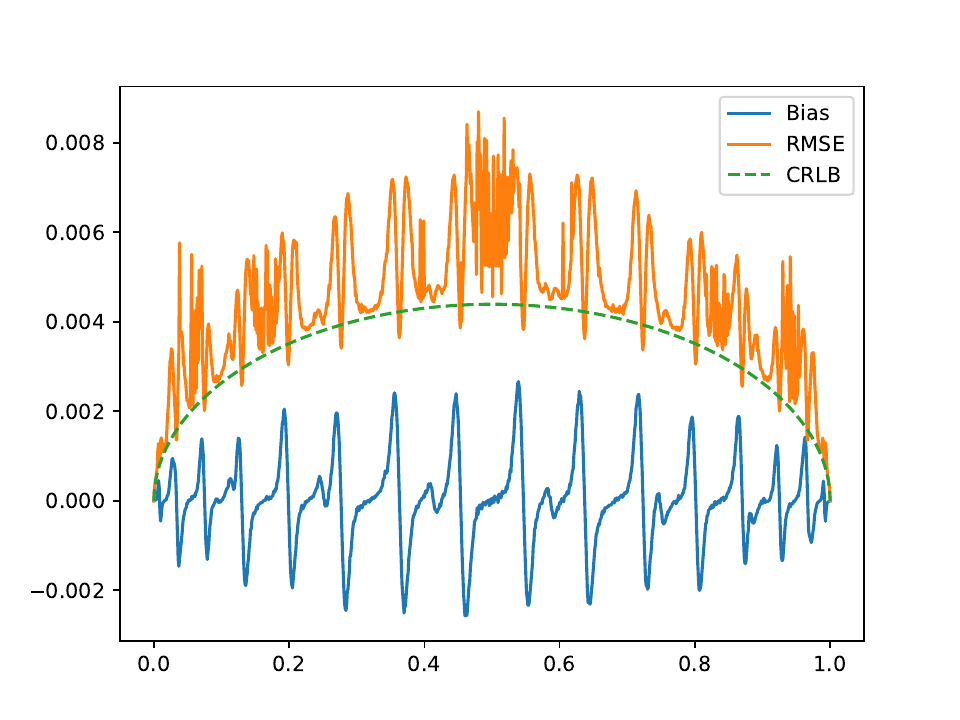}
  \caption{The bias and the root of mean squared error (RMSE) of MLAE, for different $a$. The unbiased Cramér-Rao lower bound (CRLB) is the ideal distribution of RMSE, which is equal to \autoref{eq:cramer-rao} where $b(a)=0$.}
  \label{fig:mlae-intro}
\end{figure}

In this paper, we dive further into MLAE.
In more precise experiments we find that the MLAE algorithm is not unbiased, and the bias behaves periodically with respect to the ground truth $a$, as shown in \autoref{fig:mlae-intro}.
Moreover, statistics theories show that the variance of any estimation $\tilde{a}$ follows the \textit{Cramér-Rao inequality}~\cite{CramerRao},
\begin{equation}
  \mathbb{E}[(\tilde{a}-a)^2] \geq \frac{[1+b'(a)]^2}{\mathcal{F}(a)}+b(a)^2,
  \label{eq:cramer-rao}
\end{equation}
where $b(a)=\mathbb{E}[\tilde{a}-a]$ is the \textit{bias}, and the \textit{Fisher information} $\mathcal{F}$ is defined as,
\begin{equation}
    \mathcal{F}(a) = \mathbb{E}\left[\left(\frac{\partial \ln L(a)}{\partial a}\right)^2\right],
\end{equation}
where $L$ is the likelihood function of MLAE.
An estimation is \textit{fully efficient}~\cite{fisher1922mathematical} if it is unbiased and saturates the Cramér-Rao inequality.
From \autoref{fig:mlae-intro}, we can see that MLAE is approximately unbiased and close to the unbiased Cramér-Rao lower bound in most area, except some periodical small intervals.
We propose a qualitative theory for this phenomenon, and based on the theory we propose a \textit{Random-depth Quantum Amplitude Estimation} (RQAE) algorithm.

The contributions of this paper are,
\begin{itemize}
  \item Propose a qualitative theory for the bias of MLAE that repeated use of the same amplitude amplification depths $M$ will cause a strong bias around the so-called \textit{critical points of order $M$};
  \item Introduce the implementation of even-depth amplitude amplifications, while only odd-depth ones are considered in history research;
  \item Propose a \textit{Random-depth Quantum Amplitude Estimation} (RQAE) algorithm based on the critical point theory, and use numerical experiments to show that RQAE has lower error level compared to MLAE and some other algorithms, and approaches the Heisenberg limit at about $N\cdot\epsilon=2.7\sim 2.9$.
\end{itemize}

\section{Preliminary}\label{sec:preliminary}

Most amplitude estimation algorithms are based on a general procedure called \textit{amplitude amplification}~\cite{brassard2002quantum}, which performs the transformation
\begin{align}
    & \mathcal{Q}^m\mathcal{A}\ket{00\cdots0} \nonumber
    \\ = & \cos[(2m+1)\phi]\ket{\psi_0}\ket{0} + \sin[(2m+1)\phi]\ket{\psi_1}\ket{1},
    \label{eq:aa-odd}
\end{align}
where
\begin{equation}
    \mathcal{Q} = \mathcal{A}(2\dyad{00\cdots0}-\mathbf{I})\mathcal{A}^{-1}(\mathbf{I}\otimes\mathbf{Z}).
\end{equation}

By measuring the last qubit with respect to the computational basis we obtain one with probability $\sin^2[(2m+1)\phi]$, and zero with probability $\cos^2[(2m+1)\phi]$.
Such amplitude amplification process requires $(2m+1)$ calls to the oracle $\mathcal{A}$.

Here we introduce the MLAE algorithm for amplitude estimation.
In MLAE we run a sequence of amplitude amplifications with different depths $\{m_k\}$ and number of repetitions $\{R_k\}$ for $k=1,2,\cdots,K$.
For each $k$ the state $\mathcal{Q}^{m_k}\mathcal{A}\ket{00\cdots0}$ is measured for $R_k$ times.
Let $h_k$ be the number of ones in all $R_k$ measurement results.
The final estimation $\tilde{a}$ is obtained by maximizing the likelihood function
\begin{equation}
    L(a) := \prod_{k=1}^K \ell_k(\phi),
    \label{eq:mle}
\end{equation}
where $a\equiv\sin^2\phi$, and
\begin{equation}
    \ell_k(\phi) := \left[\sin^2(M_k\phi)\right]^{h_k} \left[\cos^2(M_k\phi)\right]^{R_k-h_k},
\end{equation}
where $M_k\equiv 2m_k+1$ is called the \textit{depth} in the paper.

\begin{figure}
    \centering
    \includegraphics[width=.48\textwidth]{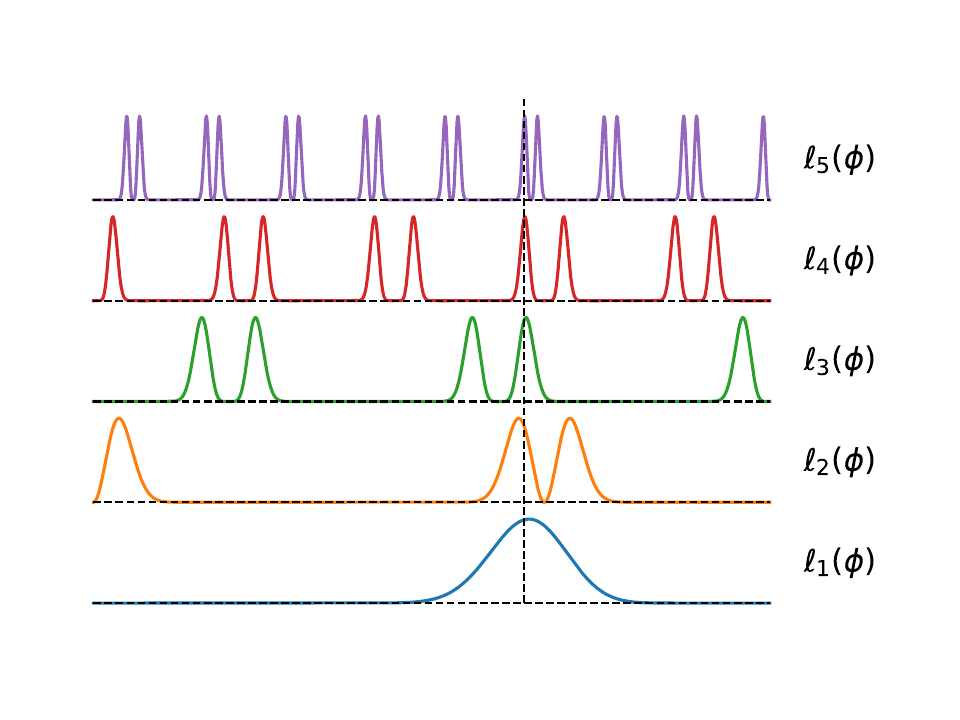}
    \caption{An illustration of how MLAE works. The curves illustrate the function $\ell_k(\phi)$ for each $k$. Here $M_1=1$, $M_k=2^k+1(k=2,3,4,5)$. }
    \label{fig:illu-mlae}
\end{figure}

The original article about MLAE algorithm~\cite{suzuki2020amplitude} presents two strategies of choosing parameters,
\begin{itemize}
    \item Linear Incremental Sequence (LIS): $m_k=k-1$ and $R_k=R$ for $k=1,2,\cdots,K$, which has error convergence $\varepsilon\sim N^{-3/4}$;
    \item Exponential Incremental Sequence (EIS): $m_1=0$, $m_k=2^{k-2}(k=2,3,\cdots,K)$ and $R_k=R(k=1,2,\cdots,K)$, which has error convergence $\varepsilon\sim N^{-1}$.
\end{itemize}

We write the MLAE algorithm in \autoref{alg:MLAE}, and illustrate how MLAE works in \autoref{fig:illu-mlae}.
Generally the function $\ell_k(\phi)$ has $M_k$ peaks.
For $M_1=1$, there is a single smooth peak in the likelihood function $\ell_1(\phi)$.
For bigger $M_k$s, the peaks are sharper and thus have better estimation ability, but there are more than one peaks.
So we cannot get more accurate estimation with $\ell_k(\phi)$ alone.
The MLAE algorithm combines the information of $\ell_k(\phi)$ for different $M_k$s by multiplying all those likelihood functions, thus obtaining a likelihood function $L$ that has only one sharp peak.

\begin{algorithm}
    \SetKwInOut{Input}{Input}
    \SetKwInOut{Output}{Output}
    \Input{ $K$: Number of iterations; $R$: Number of measurements in each iteration; }
    \Output{ $\tilde{a}$: Estimation of $a$; }

    \For{$k=1..K$}{
        Set $m_k=2^{k-1}$ and $R_k=R$;

        Measure the last qubit of $\mathcal{Q}^{m_k}\mathcal{A}\ket{00\cdots0}$ for $R_k$ times and let $h_k$ be the number of ones;
    }

    Calculate $\tilde{a}$ using MLE in \autoref{eq:mle}.

    \caption{Maximum Likelihood Amplitude Estimation (MLAE) with Exponential Incremental Sequence (EIS)}
    \label{alg:MLAE}
\end{algorithm}

By calculation the Fisher information of MLAE is~\cite{suzuki2020amplitude},
\begin{equation}
    \mathcal{F}(a) = \frac{1}{a(1-a)}\sum_k R_kM_k^2.
    \label{eq:fisher}
\end{equation}

In most application problems the major complexity lies in the oracle $\mathcal{A}$ itself.
Therefore, the time cost of MLAE is,
\begin{equation}
    N = \sum_k R_kM_k.
\end{equation}

As MLAE is approximately unbiased and saturates the Cramér-Rao inequality in most area, the RMSE has the same error convergence as $\mathcal{F}^{-1/2}$.
The MLAE algorithm with EIS fixes $R_1=\cdots=R_K=R$, and chooses $M_1=1,M_k=2^{k-1}+1(k\ge2)$, then $N=O(R\cdot 2^K)$ and $\mathcal{F}^{-1/2}=O(R^{-1/2}\cdot 2^{-K})=O(N^{-1})$, which reaches the Heisenberg limit asymptotically and is quadratically faster than MC.
But as is shown in our experiments, if we simply fix the $R$ then the difference between RMSE and Cramér-Rao lower bound~(CRLB) grows, which makes MLAE error convergence slower than the Heisenberg limit.

\section{Theory}\label{sec:theory}

\subsection{Critical Point}\label{subsec:c-point}

\begin{figure*}
    \centering
    \subfloat[$R=32$,$\{M_k\}=\{1,3,5,9,17\}$.]{
        \includegraphics[width=.48\textwidth]{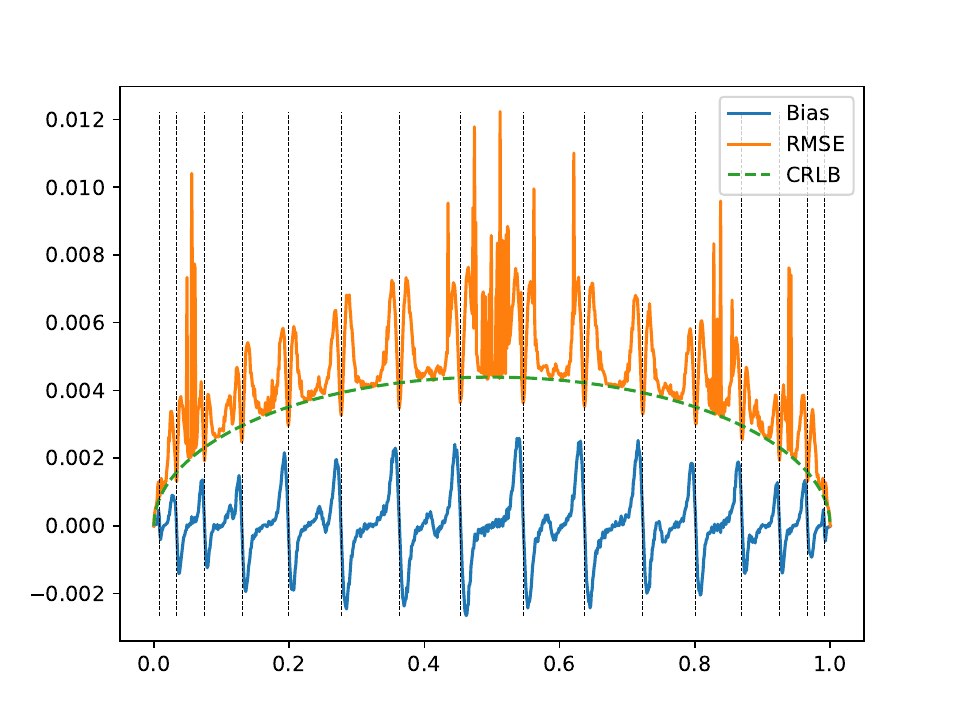}
        \label{fig:bias-mlae-17}
    }
    \subfloat[$R=32$,$\{M_k\}=\{1,3,5,9,18\}$.]{
        \includegraphics[width=.48\textwidth]{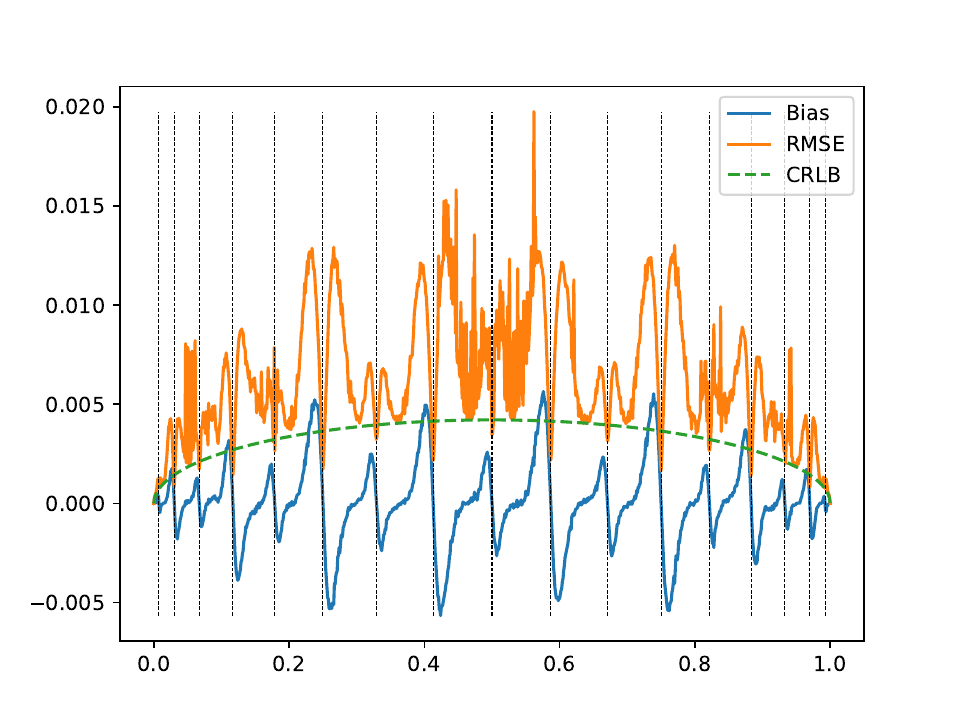}
        \label{fig:bias-mlae-18}
    }

    \subfloat[$R=32$,$\{M_k\}=\{1,3,5,9,19\}$.]{
        \includegraphics[width=.48\textwidth]{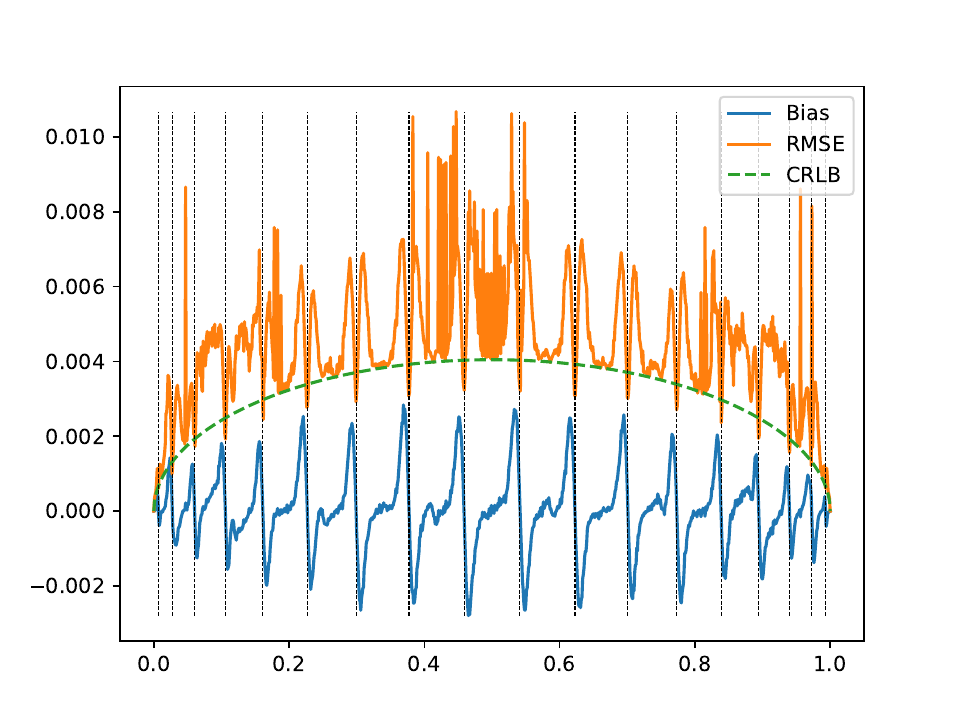}
        \label{fig:bias-mlae-19}
    }
    \subfloat[$R=32$,$\{M_k\}=\{1,2,4,8,16\}$.]{
        \includegraphics[width=.48\textwidth]{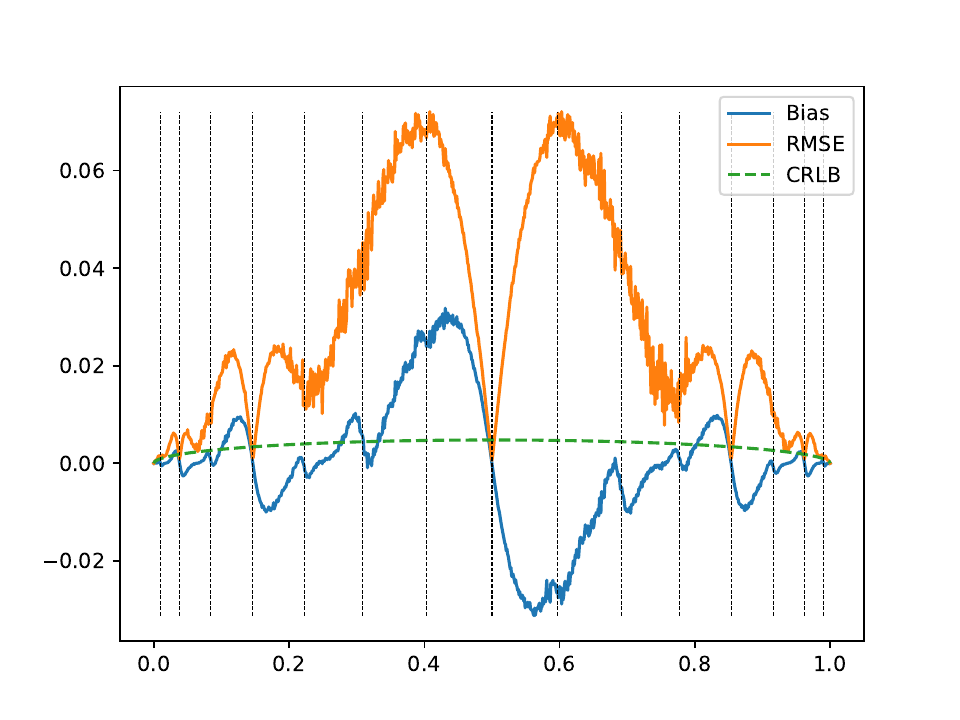}
        \label{fig:bias-mlae-extreme}
    }
    \caption{ The performance of MLAE with critical points of order $M_K$ labelled by vertical dashed lines ($x$-axis: the amplitude $a$, $y$-axis: bias, RMSE or CRLB). stan For each $a$ we simulate 4096 times to calculate the bias and RMSE. \autoref{sub@fig:bias-mlae-17}~-\autoref{sub@fig:bias-mlae-19} A comparison among 3 configurations with similar parameter choices. The parameter choice in \autoref{sub@fig:bias-mlae-17} is the EIS choice with $K=5$ and $R=32$. \autoref{sub@fig:bias-mlae-extreme} An extreme case, in which the error level is much higher than other three cases. }
    \label{fig:bias-mlae-coprime}
\end{figure*}

In the beginning of this section, we set up a model for the bias of MLAE.
In MLAE, consider the two values $a_\pm=\sin^2(\phi_\pm)=\sin^2(j\pi/2M_K\pm\delta)$ for small $\delta$.
The depth-$M_K$ amplitude amplification cannot tell apart $a_\pm=\sin^2(\phi_\pm)=\sin^2(j\pi/2M_K\pm\delta)$, as $\ell_K(\phi_+)=\ell_K(\phi_-)$.
Therefore, they can only be told apart by other terms $\{\ell_k(\phi)\}_{k=1}^{K-1}$ in the likelihood function that is not so sharp as $\ell_K(\phi)$.
As a result, MLAE has a positive bias when $a_-$ is the ground truth, and has a negative bias when $a_+$ is the ground truth.
We call,
\begin{equation}
  \left\{
    \sin^2\left(\frac{j}{m}\cdot\frac{\pi}{2}\right)
  \middle|
    j=1,2,\cdots,m-1
  \right\}
\end{equation}
the \textit{critical points} of order $m$.
The \textit{exceptional points} proposed in~\cite{callison2022improved} are exactly the critical points of order $M_K=\max\{M_k\}$.
The critical point theory concludes that the original MLAE algorithm has obvious bias in the intervals centered at each critical point of order $M_K$, as shown in \autoref{sub@fig:bias-mlae-17}.
All the quantum outputs in the experiments are obtained by sampling the theoretic distribution functions.
The vertical dashed lines are the critical points of order $M_K$.
The most intensive bias occurs around each dashed line, which is positive on the left of each line and negative on the right.

Furthermore, the critical points of different orders may overlap, which may lead to even more intensive bias.
In \autoref{fig:bias-mlae-17}~-\autoref{fig:bias-mlae-19}, we do numerical experiments for MLAE with 3 similar parameter choices.
We find that the bias and RMSE of \autoref{fig:bias-mlae-18} is obviously more intensive than \autoref{fig:bias-mlae-17} and \autoref{fig:bias-mlae-19}, especially in the vicinity of the common critical points of orders 3, 9 and 18.
In \autoref{sub@fig:bias-mlae-extreme}, we consider an extreme case where all $M_k$s are powers of two, $a=0.5$ is a common critical point of order 2,4,8 and 16, so the bias and RMSE behaves extremely badly.

In summary, the distribution of critical points has a significant impact on the error behavior of MLAE.
Usually, the most intensive bias and RMSE occurs around the critical points of order $\max\{M_k\}$.
When a critical point of different orders including $\max\{M_k\}$ overlap, the bias and RMSE become even bigger.
This theory inspires us that an important task to improve the robustness of MLAE is to avoid the critical points by optimizing the parameter choices.

\subsection{The implementation of even-depth amplitude amplification}\label{subsec:aa-even}

\autoref{eq:aa-odd} enables us to generate a 0-1 distribution random variable with $p(1)=\sin^2[M\phi]$ for any odd number $M$.
But in the last subsection, our numerical experiments allow the depth $M_k$ to be even.
In this subsection we complete the theory of amplitude amplification by introducing the implementation of even-depth quantum amplitude amplification.

From \autoref{eq:def-A} we have,
\begin{equation}
    \cos\phi\mathcal{A}^{-1}(\ket{\psi_0}\ket{0}) + \sin\phi\mathcal{A}^{-1}(\ket{\psi_1}\ket{1}) = \ket{00\cdots0}.
\end{equation}

By the orthogonality of $\mathcal{A}^{-1}$ we know that,
\begin{equation}
    \ket{\psi'} := \sin\phi\mathcal{A}^{-1}(\ket{\psi_0}\ket{0}) - \cos\phi\mathcal{A}^{-1}(\ket{\psi_1}\ket{1}),
\end{equation}
is orthogonal to $\ket{00\cdots0}$.
That is, if we measure all qubits of $\ket{\psi'}$ under the computational basis, we will certainly get results that contain one.
Moreover,
\begin{eqnarray}
    \mathcal{A}^{-1}\ket{\psi_0}\ket{0} = \cos\phi\ket{00\cdots0} + \sin\phi\ket{\psi'}, \\
    \mathcal{A}^{-1}\ket{\psi_1}\ket{1} = \sin\phi\ket{00\cdots0} - \cos\phi\ket{\psi'}.
\end{eqnarray}

Define,
\begin{equation}
    \mathcal{Q}'
    =
    \mathcal{A}^{-1}(\mathbf{I}\otimes\mathbf{Z})\mathcal{A}(2\dyad{00\cdots0}-\mathbf{I}).
\end{equation}

Then,
\begin{align}
  & \mathcal{Q}'\ket{00\cdots0}
  \nonumber
  \\ = &
  \mathcal{A}^{-1}
  (\mathbf{I}\otimes\mathbf{Z})
  \mathcal{A}
  \ket{00\cdots0}
  \nonumber
  \\ = &
  \mathcal{A}^{-1}
  (\mathbf{I}\otimes\mathbf{Z})
  (\cos\phi\ket{\psi_0}\ket{0} + \sin\phi\ket{\psi_1}\ket{1})
  \nonumber
  \\ = &
  \mathcal{A}^{-1}
  (\cos\phi\ket{\psi_0}\ket{0} - \sin\phi\ket{\psi_1}\ket{1})
  \nonumber
  \\ = &
  \cos\phi(\cos\phi\ket{00\cdots0} + \sin\phi\ket{\psi'})
  \nonumber
  \\ & -\sin\phi(\sin\phi\ket{00\cdots0} - \cos\phi\ket{\psi'})
  \nonumber
  \\ = &
  \cos(2\phi)\ket{00\cdots0} + \sin(2\phi)\ket{\psi'},
\end{align}
and,
\begin{align}
    & \mathcal{Q}'\ket{\psi'}
    \nonumber
    \\ = &
    \mathcal{A}^{-1}
    (\mathbf{I}\otimes\mathbf{Z})
    \mathcal{A}
    (-\ket{\psi'})
    \nonumber
    \\ = &
    \mathcal{A}^{-1}
    (\mathbf{I}\otimes\mathbf{Z})
    (-\sin\phi\ket{\psi_0}\ket{0} + \cos\phi\ket{\psi_1}\ket{1})
    \nonumber
    \\ = &
    \mathcal{A}^{-1}
    (-\sin\phi\ket{\psi_0}\ket{0} - \cos\phi\ket{\psi_1}\ket{1})
    \nonumber
    \\ = &
    -\sin\phi(\cos\phi\ket{00\cdots0} + \sin\phi\ket{\psi'})
    \nonumber
    \\ &-\cos\phi(\sin\phi\ket{00\cdots0} - \cos\phi\ket{\psi'})
    \nonumber
    \\ = &
    -\sin(2\phi)\ket{00\cdots0} + \cos(2\phi)\ket{\psi'}.
\end{align}

Therefore, $\mathcal{Q}'$ is a rotation by angle $2\phi$ in the plane spanned by $\ket{00\cdots0}$ and $\ket{\psi'}$.
We can deduce that,
\begin{equation}
    \mathcal{Q}'^m\ket{00\cdots0} = \cos(2m\phi)\ket{00\cdots0} + \sin(2m\phi)\ket{\psi'}.
\end{equation}

By measuring all qubits under the computational basis we obtain all \textit{zero} with probability $\cos^2(2m\phi)$, and results containing at least one \textit{one} with probability $\sin^2(2m\phi)$.
The extended amplitude amplification process requires $2m$ calls to the oracle $\mathcal{A}$.

In summary, no matter $M$ is odd or even, we can obtain a random variable $r_M$ with 0-1 distribution where $p(1)=\sin^2(M\phi)$, with a cost of $M$ oracle calls to the oracle $\mathcal{A}$.
When $M$ is odd, we measure the last qubit of the state $\mathcal{Q}^{(M-1)/2}\mathcal{A}\ket{00\cdots0}$, and obtain one with probability $\sin^2(M\phi)$.
When $M$ is even, we measure all qubits of the state $\mathcal{Q}'^{M/2}\ket{00\cdots0}$, and the probability that the results contain one is $\sin^2(M\phi)$.
For convenience, we use the terminology \textit{measuring $r_M$} to mean that we use the procedure above to obtain a random variable of 0-1 distribution with $p(1)=\sin^2(M\phi)$.
The extended amplitude amplification is crucial to our proposed algorithm.

\section{Algorithm}\label{sec:alg}

\begin{figure*}
  \centering
  \subfloat[DJQAE with $K=6$ and $R=32$.]{
    \includegraphics[width=.48\textwidth]{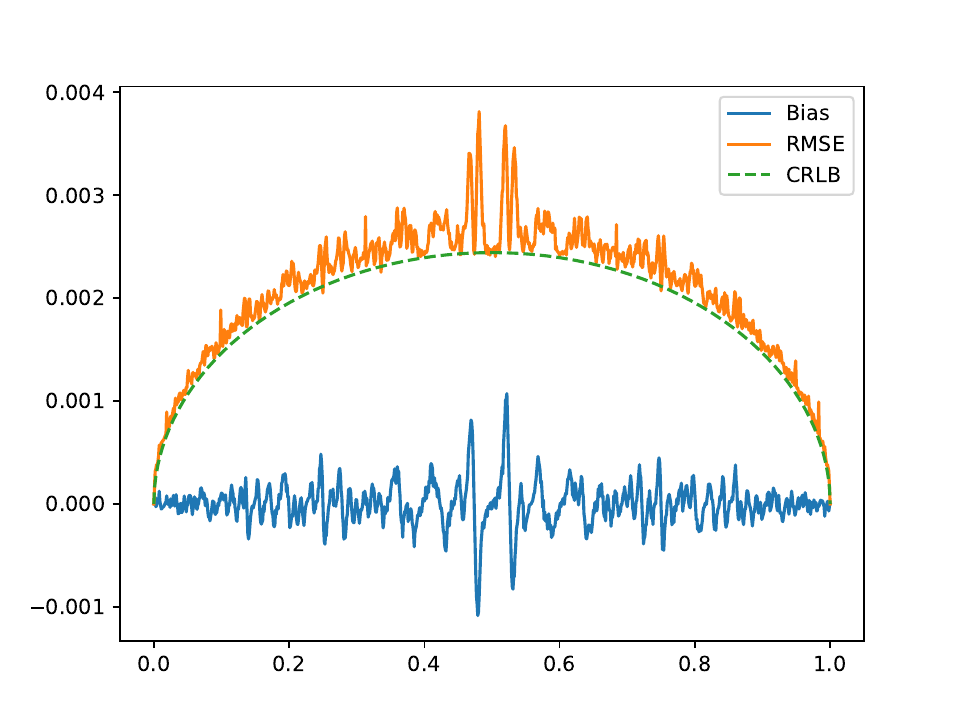}
    \label{fig:bias-djqae}
  }
  \subfloat[RQAE with $K=5$ and $R=32$.]{
    \includegraphics[width=.48\textwidth]{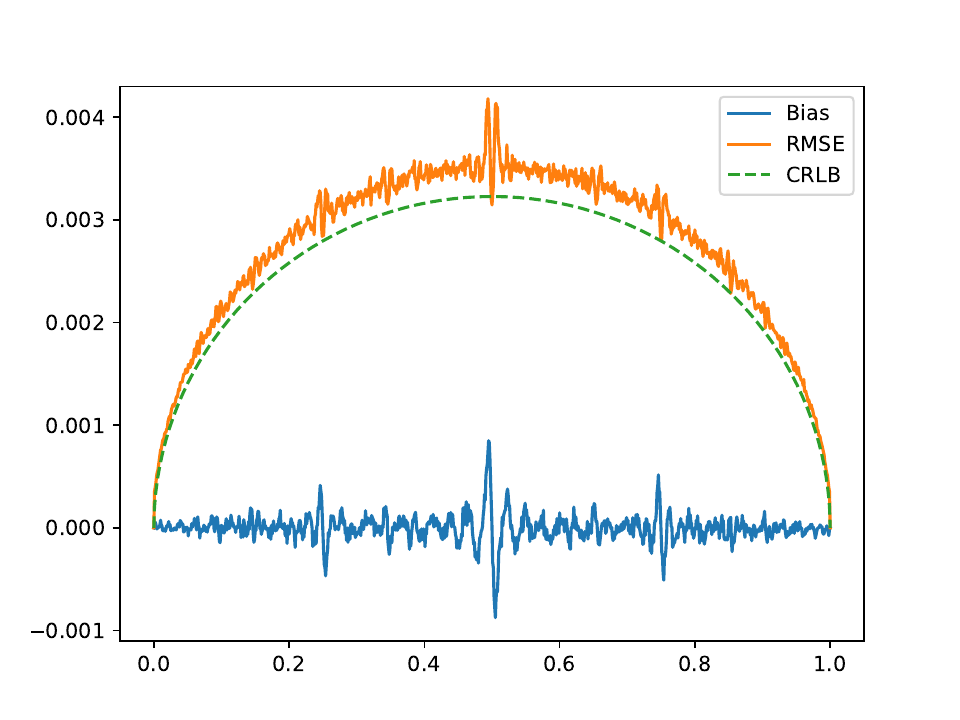}
    \label{fig:bias-rqae}
  }
  \caption{The performance of DJQAE and RQAE ($x$-axis: the amplitude $a$, $y$-axis: bias, RMSE or CRLB). For each $a$ we simulate 4096 times to calculate the bias and RMSE. }
\end{figure*}

\begin{algorithm}[t]
  \SetKwInOut{Input}{Input}
  \SetKwInOut{Output}{Output}
  \Input{ $K$: Number of iterations; $R$: Number of measurements in each iteration; }
  \Output{ $\tilde{a}$: Estimation of $a$; }

  Set $M_1=1$ and $R_1=R$;

  \For{$i=2..K$}{
    Set $M_m=m$ and $R_m=0$ for $m=2^{i-1},\cdots,2^{i}-1$;

    Draw $R$ random samples from $\{2^{i-1},\cdots,2^{i}-1\}$ with equal probabilities;
  }

  \For{$m=1..2^K-1$}{
    \If{$R_m>0$}{
      \If{$M_m$ is even}{
        Measure all qubits of $\mathcal{Q}^{\prime M_m/2}\ket{00\cdots0}$ for $R_m$ times and let $h_m$ be the number of results that are not all zero;
      }
      \Else{
        Measure the last qubit of $\mathcal{Q}^{\prime (M_m-1)/2}\mathcal{A}\ket{00\cdots0}$ for $R_m$ times and let $h_m$ be the number of ones;
      }
    }
  }

  Calculate $\tilde{a}$ using maximum likelihood estimation.

  \caption{Random-depth Quantum Amplitude Estimation (RQAE)}
  \label{alg:RQAE}
\end{algorithm}

As is discussed in the critical point theory in the last section, the key idea to improve MLAE is to avoid critical points.
The depth-jittering quantum amplitude estimation (DJQAE)~\cite{callison2022improved} avoids the critical points of order $M_k$ by jittering the depth $M_k$ into a range $\{M_k^{(L)},\cdots,M_k^{(U)}\}$, and the number of repetitions of each depth in that range is summed to $R$.
Since the number of repetitions of each depth is reduced, the impact brought by critical points of the corresponding orders is also reduced.
As shown in \autoref{fig:bias-djqae}, the jittering step can reduce the intensity of the bias and RMSE remarkably, but still far from unbiasedness.

The critical point theory concludes that too many repeated use of the same $M$ will cause bad behavior around critical points of order $M$.
The intuition of our RQAE algorithm is that we can randomly choose the next depths to reduce the impact of the critical points of a specific order.
We improve MLAE with EIS and propose the RQAE algorithm, as presented in \autoref{alg:RQAE}.
A simple demonstration that RQAE successfully reduces the impact of critical points is shown in \autoref{fig:bias-rqae}.

The expected number of queries $N$ and Fisher information $\mathcal{F}$ are,
\begin{equation}
  N = \sum_{k=0}^{K} \frac{R}{2^k}\sum_{M=2^k}^{2^{k+1}-1}M
    = \Theta(2^KR),
\end{equation}
\begin{equation}
  \mathcal{F} = \sum_{k=0}^{K} \frac{R}{2^k}\sum_{M=2^k}^{2^{k+1}-1}M^2
  = \Theta(4^KR),
\end{equation}
so conditioned on fixed $R$, the RMSE is asymptotically $\mathcal{F}^{-1/2} = \Theta(N^{-1})$.
This guarantees that RQAE asymptotically reaches the Heisenberg limit as $K\rightarrow\infty$, but for finite $K$ the CRLB is usually not saturated, and it remains to the experiments to show how well the Heisenberg limit is approached.

\section{Experiments}\label{sec:expr}

\begin{figure*}
    \centering
    \subfloat[MLAE]{
      \includegraphics[width=.98\textwidth]{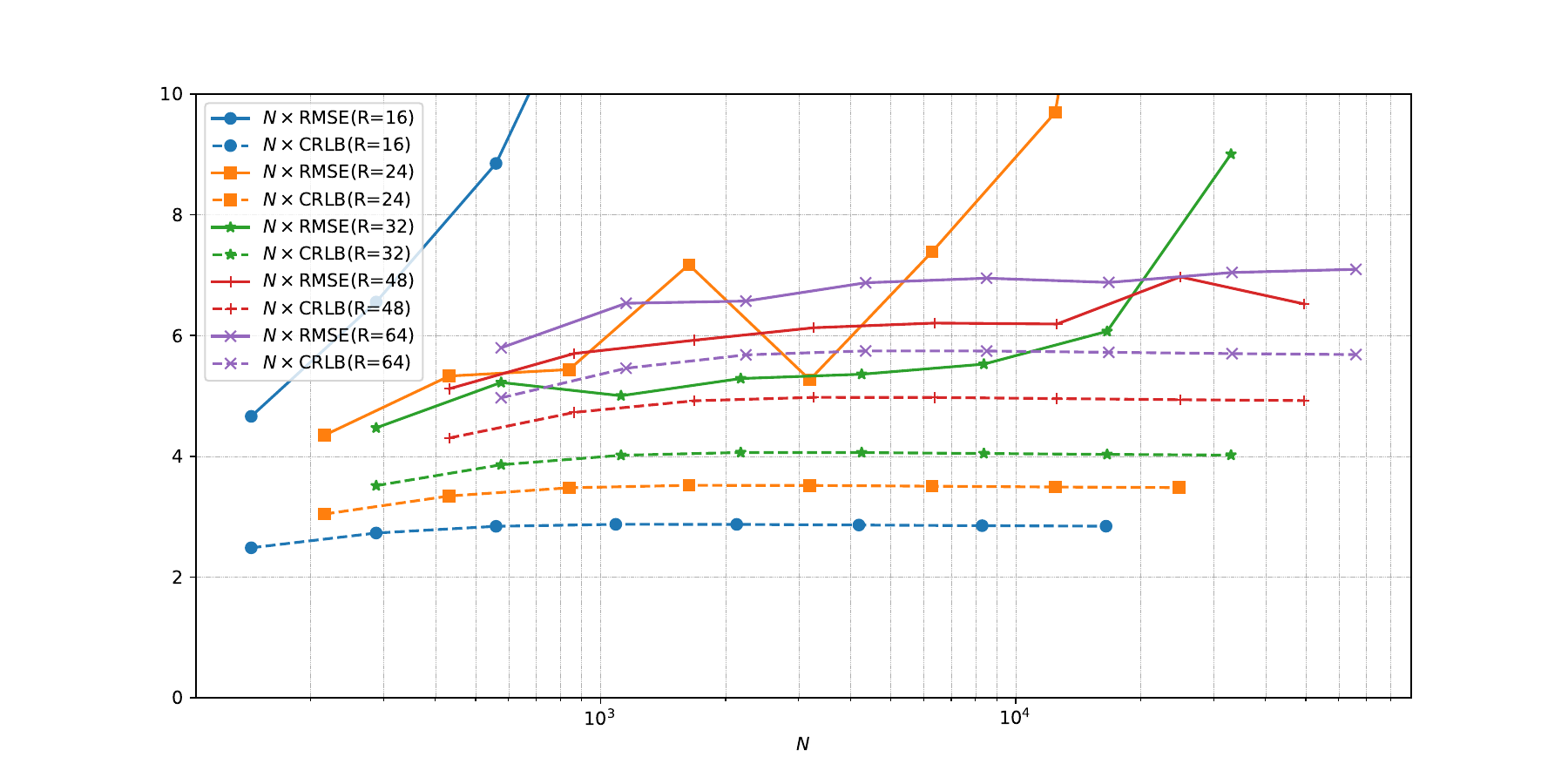}
      \label{fig:param-mlae}
    }\\
    \subfloat[RQAE]{
      \includegraphics[width=.98\textwidth]{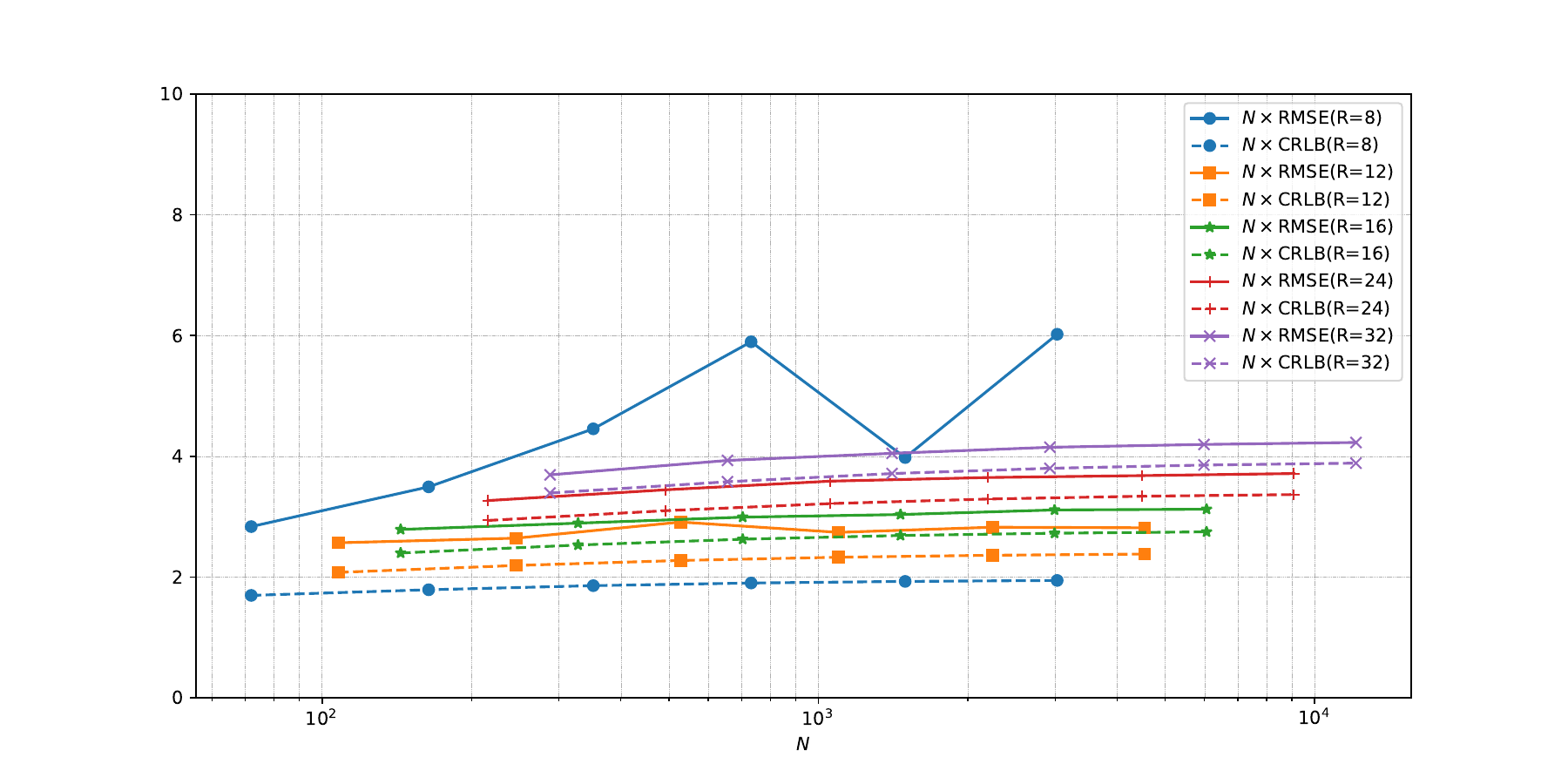}
      \label{fig:param-rqae}
    }
    \caption{The RMSE and CRLB curve for different $K$ and $R$ ($x$-axis: the number of oracle calls $N$, $y$-axis: RMSE or CRLB). For each curve we fix the $R$ and let $K$. In each figure, the solid and dashed curves of the same color stand for the RMSE and CRLB curve with the same configurations. For each configuration we simulate 65536 times to calculate the RMSE and CRLB.}
\end{figure*}

\begin{figure*}
  \centering
  \includegraphics[width=\textwidth]{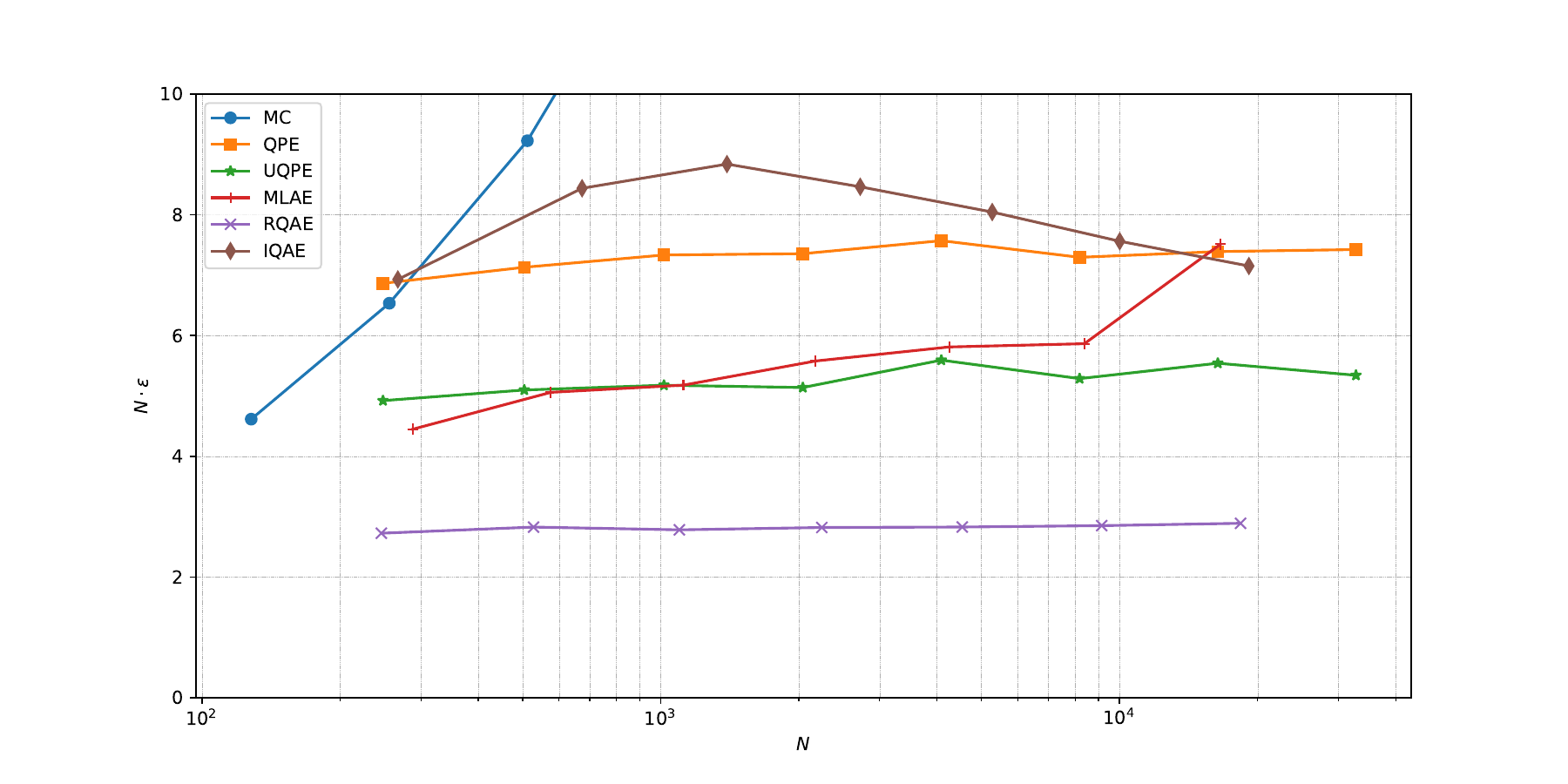}
  \caption{The error behavior $\epsilon$ with respect to the time cost $N$. Note that we use $N\cdot\epsilon$ on the $y$-axis to demonstrate the Heisenberg scaling. }
  \label{fig:comp-rd}
\end{figure*}

In this section, we use numerical experiments to guide the parameter setting of RQAE and then compare its performance with other algorithms.

By reducing the impact of critical points, the RMSE curve of RQAE is expected to be closer to CRLB than that of MLAE.
We prove this result with numerical experiments in \autoref{fig:param-mlae} and \autoref{fig:param-rqae}.
In both figures, the CRLB curves grow horizentally as $N$ increases, which mean a nice Heisenberg scaling.
But the gap between each pair of RMSE and CRLB curves also plays an important role.
In \autoref{fig:param-mlae}, we can see that each pair of RMSE curve and the corresponding CRLB is close to each other when $K$ is small, and the RMSE curve gets suddenly away from the CRLB when $K$ reaches a certain threshold.
The larger $R$ is, the larger the threshold is.
As a result, the behavior of MLAE may not reach the Heisenberg limit since one need to fix $R$ to reach the Heisenberg limit, as deducted in \autoref{sec:preliminary}.
On the other hand, in \autoref{fig:param-rqae}, the RMSE curve of RQAE is only away from the CRLB when $R$ is small.
For merely $R\geq 12$, the gap becomes quite small, which makes RQAE able to reach better Heisenberg limit.

We compare different amplitude estimation algorithms and take the time cost into consideration.
In this experiment we uniformly randomly draw $2^{20}$ samples in the interval $[0,1]$ as $a$, and compare the error behavior with respect to the time cost.
For Monte Carlo~(MC) estimation, suppose the state \autoref{eq:def-A} is prepared for $R$ times, and by measuring the last qubit the result \textit{one} is obtained for $h$ times, then the estimation to $a$ is given by $\tilde{a}=h/R$.
The time cost for MC is $N=R$, as each preparation of the state \autoref{eq:def-A} requires one call to the oracle $\mathcal{A}$.
For MLAE and RQAE, we fix $R=32$ and $R=12$ respectively, and let $K$ vary.
The \textit{quantum phase estimation}~(QPE) based amplitude estimation estimates the rotation angle $2\phi$, as is described in \autoref{subsec:aa-even}.
Since one rotation requires two calls to the oracle $\mathcal{A}$, the time cost for QPE is $N=2\sum_{j=0}^{t-1}2^j=2(2^t-1)$.
An efficient way to reduce the RMSE of QPE is to repeat for $R$ times and use MLE to give the final estimation.
The \textit{unbiased quantum phase estimation}~(UQPE)~\cite{lu2023unbiased} is an unbiased variant of QPE.
The time cost for both QPE and UQPE in our experiments is $N=2R(2^t-1)$, where the notations come from \cite{lu2023unbiased}.
In our experiments we fix $R=4$ and let $t$ vary.
For IQAE~\cite{Grinko2021}, we use Chernoff-Hoeffding confidence interval method, fix $\alpha=0.05,N_{\text{shots}}=32$ and let $\epsilon$ vary.
The results are shown in \autoref{fig:comp-rd}.
The MC algorithm have an error convergence of $O(N^{-1/2})$, while all other algorithms have an asymptotic $O(N^{-1})$ error convergence.
By choosing nearly optimized parameter set for each algorithm, we can see that the RQAE algorithm outperforms all other algorithms and approaches the Heisenberg limit at about $N\cdot\epsilon=2.7\sim 2.9$.

\section{Conclusion}\label{sec:conclusion}

The maximum likelihood amplitude estimation (MLAE) algorithm is a practical solution to the quantum amplitude estimation problem, which has a theoretically quadratic speedup over classical Monte Carlo method.
We find that MLAE behaves efficient, i.e., unbiased and saturates the Cramér-Rao inequality in most area except some periodical small intervals.
We set up a critical point model and analyze how the bias is influenced by the distribution of the critical points.
Also, we introduce the implementation of even-depth quantum amplitude amplification.
Then, we propose a \textit{Random-depth Quantum Amplitude Estimation} (RQAE) algorithm by choosing MLAE depths randomly to reduce the impact of critical points of a specific order.
In the end, we do numerical experiments among some amplitude estimation algorithms, including Monte Carlo estimation, quantum phase estimation and its unbiased variant, iterative quantum amplitude estimation, maximum likelihood amplitude estimation, depth-jittering quantum amplitude estimation and our random-depth quantum amplitude estimation.
We show that our algorithm performs the best among all algorithms and approaches the Heisenberg limit at about $N\cdot\epsilon=2.7\sim 2.9$.


\printbibliography

\end{document}